%% file: ms.tex
\documentclass[]{aastex631}

\usepackage{float}
\usepackage{amsmath}
\usepackage{tasks}
\usepackage{booktabs}
\usepackage{xspace}

\newcommand{\projname}{DESI Strong Lens Foundry\xspace}

\newcommand{\ang}{\AA\xspace}
\newcommand{\hst}{\emph{HST}\xspace}
\newcommand{\HST}{\emph{Hubble Space Telescope}\xspace}

\newcommand\pound{\scalebox{0.8}{\raisebox{0.4ex}{\#}}}
\newcommand{\ed}{\textcolor{black}}
\newcommand{\desied}{\textcolor{black}}
\newcommand{\apjed}{\textcolor{black}}

\shorttitle{\projname III}
\shortauthors{Agarwal, Huang et al.}

\graphicspath{{./}{figures/}}

\begin{document}

\title{\projname III:\\
Keck Spectroscopy for Strong Lenses Discovered Using Residual Neural Networks
}

\correspondingauthor{Shrihan Agarwal, Xiaosheng Huang}
\email{shrihan@uchicago.edu, xhuang22@usfca.edu}

\author[0000-0002-2350-4610]{Shrihan~Agarwal}
\affiliation{University of Chicago, Department of Astronomy, Chicago, IL 60615, USA}
\affiliation{University of California, Berkeley, Department of Astronomy, Berkeley, CA 94720, USA}
\affiliation{Physics Division, Lawrence Berkeley National Laboratory, 1 Cyclotron Road, Berkeley, CA 94720, USA}

\author[0000-0001-8156-0330]{Xiaosheng~Huang}
\affiliation{Department of Physics \& Astronomy, University of San Francisco, San Francisco, CA 94117, USA}
\affiliation{Physics Division, Lawrence Berkeley National Laboratory, 1 Cyclotron Road, Berkeley, CA 94720, USA}

\author[0000-0003-1889-0227]{W.~Sheu}
\affiliation{Physics Division, Lawrence Berkeley National Laboratory, 1 Cyclotron Road, Berkeley, CA 94720, USA}
\affiliation{Department of Physics \& Astronomy, University of California, Los Angeles, Los Angeles, CA 90095, USA}

\author[0000-0002-0385-0014]{C.J.~Storfer}
\affiliation{Institute for Astronomy, University of Hawai'i, Honolulu, HI 96822-1897, USA}

\author[0009-0008-0518-8045]{M. ~Tamargo-Arizmendi}
\affiliation{Department of Physics and Astronomy and PITT PACC, University of Pittsburgh, Pittsburgh, PA 15260, USA}

\author[0000-0003-3533-5890]{S. Tabares-Tarquinio}
\affiliation{Physics Division, Lawrence Berkeley National Laboratory, 1 Cyclotron Road, Berkeley, CA 94720, USA}
\affiliation{Department of Physics \& Astronomy, University of San Francisco, San Francisco, CA 94117, USA}

\author[0000-0002-5042-5088]{D.J.~Schlegel}
\affiliation{Physics Division, Lawrence Berkeley National Laboratory, 1 Cyclotron Road, Berkeley, CA, 94720
}

\author{G. Aldering}
\affiliation{Physics Division, Lawrence Berkeley National Laboratory, 1 Cyclotron Road, Berkeley, CA 94720, USA}

\author[0000-0002-4928-4003]{A.~Bolton}
\affiliation{NSF's National Optical-Infrared Astronomy Research Laboratory, 950 N. Cherry Ave., Tucson, AZ 85719}

\author[0000-0001-7101-9831]{A.~Cikota}
\affiliation{Gemini Observatory / NSF's NOIRLab, Casilla 603, La Serena, Chile}

\author[0000-0002-4928-4003]{Arjun~Dey}
\affiliation{NSF's National Optical-Infrared Astronomy Research Laboratory, 950 N. Cherry Ave., Tucson, AZ 85719}

\author[0000-0003-4701-3469]{A.~Filipp}
\affiliation{Universit\'{e} de Montréal, Physics Department, 1375 Av. Th\'{e}r\`{e}se-Lavoie-Roux, H2V 0B3 Montr\'{e}al, QC, Canada}
\affiliation{Ciela -- Montreal Institute for Astrophysical Data Analysis and Machine Learning, 1375 Av. Th\'{e}r\`{e}se-Lavoie-Roux,
H2V 0B3 Montréal, QC, Canada}
\affiliation{Technical University Munich, TUM School of Natural Sciences, Physics Department, 85748 Garching, Germany}
\affiliation{Max Planck Institute for Astrophysics (MPA), Karl-Schwarzschlid-Strasse 1, 85748 Garching, Germany}
\author[0000-0002-9253-053X]{E.~Jullo}
\affiliation{Aix-Marseille Univ., CNRS, CNES, LAM, Marseille, France}

\author[0000-0001-9802-362X]{K.J.~Kwon}
\affiliation{University of California, Santa Barbara, Santa Barbara, CA 93106, USA}

\author[0000-0002-4436-4661]{S.~Perlmutter}
\affiliation{Physics Division, Lawrence Berkeley National Laboratory, 1 Cyclotron Road, Berkeley, CA 94720, USA}
\affiliation{Department of Astronomy, University of California, Berkeley, CA 94720, USA}

\author[0000-0002-9063-698X]{Y.~Shu}
\affiliation{Purple Mountain Observatory, Chinese Academy of Sciences, Nanjing 210023, People's Republic of China}

\author[0000-0002-1106-4881]{E.~Sukay}
\affiliation{Johns Hopkins University, Department of Physics \& Astronomy, Baltimore, MD 21218, USA}

\author[0000-0001-7266-930X]{N.~Suzuki}
\affiliation{Physics Division, Lawrence Berkeley National Laboratory, 1 Cyclotron Road, Berkeley, CA 94720, USA}
\affiliation{Kavli Institute for the Physics and Mathematics of the Universe, University of Tokyo Kashiwa 277-8583, Japan}

\author{J.~Aguilar}
\affiliation{Lawrence Berkeley National Laboratory, 1 Cyclotron Road, Berkeley, CA 94720, USA}

\author[0000-0001-6098-7247]{S.~Ahlen}
\affiliation{Physics Dept., Boston University, 590 Commonwealth Avenue, Boston, MA 02215, USA}

\author[0000-0001-5537-4710]{S.~BenZvi}
\affiliation{Department of Physics \& Astronomy, University of Rochester, 206 Bausch and Lomb Hall, P.O. Box 270171, Rochester, NY 14627-0171, USA}

\author{D.~Brooks}
\affiliation{Department of Physics \& Astronomy, University College London, Gower Street, London, WC1E 6BT, UK}

\author{T.~Claybaugh}
\affiliation{Lawrence Berkeley National Laboratory, 1 Cyclotron Road, Berkeley, CA 94720, USA}

\author{P.~Doel}
\affiliation{Department of Physics \& Astronomy, University College London, Gower Street, London, WC1E 6BT, UK}

\author[0000-0002-2890-3725]{J.~E.~Forero-Romero}
\affiliation{Observatorio Astron\'omico, Universidad de los Andes, Cra. 1 No. 18A-10, Edificio H, CP 111711 Bogot\'a, Colombia}
\affiliation{Departamento de F\'isica, Universidad de los Andes, Cra. 1 No. 18A-10, Edificio Ip, CP 111711, Bogot\'a, Colombia}

\author{E.~Gaztañaga}
\affiliation{Institute of Cosmology and Gravitation, University of Portsmouth, Dennis Sciama Building, Portsmouth, PO1 3FX, UK}
\affiliation{Institute of Space Sciences, ICE-CSIC, Campus UAB, Carrer de Can Magrans s/n, 08913 Bellaterra, Barcelona, Spain}
\affiliation{Institut d'Estudis Espacials de Catalunya (IEEC), c/ Esteve Terradas 1, Edifici RDIT, Campus PMT-UPC, 08860 Castelldefels, Spain}

\author[0000-0003-3142-233X]{S.~Gontcho A Gontcho}
\affiliation{Lawrence Berkeley National Laboratory, 1 Cyclotron Road, Berkeley, CA 94720, USA}

\author{G.~Gutierrez}
\affiliation{Fermi National Accelerator Laboratory, PO Box 500, Batavia, IL 60510, USA}

\author[0000-0002-6550-2023]{K.~Honscheid}
\affiliation{The Ohio State University, Columbus, 43210 OH, USA}
\affiliation{Center for Cosmology and AstroParticle Physics, The Ohio State University, 191 West Woodruff Avenue, Columbus, OH 43210, USA}
\affiliation{Department of Physics, The Ohio State University, 191 West Woodruff Avenue, Columbus, OH 43210, USA}

\author[0000-0002-6024-466X]{M.~Ishak}
\affiliation{Department of Physics, The University of Texas at Dallas, 800 W. Campbell Rd., Richardson, TX 75080, USA}

\author[0000-0002-0000-2394]{S.~Juneau}
\affiliation{NSF NOIRLab, 950 N. Cherry Ave., Tucson, AZ 85719, USA}

\author{R.~Kehoe}
\affiliation{Department of Physics, Southern Methodist University, 3215 Daniel Avenue, Dallas, TX 75275, USA}

\author[0000-0003-3510-7134]{T.~Kisner}
\affiliation{Lawrence Berkeley National Laboratory, 1 Cyclotron Road, Berkeley, CA 94720, USA}

\author[0000-0003-2644-135X]{S.~E.~Koposov}
\affiliation{Institute for Astronomy, University of Edinburgh, Royal Observatory, Blackford Hill, Edinburgh EH9 3HJ, UK}
\affiliation{Institute of Astronomy, University of Cambridge, Madingley Road, Cambridge CB3 0HA, UK}

\author{A.~Lambert}
\affiliation{Lawrence Berkeley National Laboratory, 1 Cyclotron Road, Berkeley, CA 94720, USA}

\author[0000-0003-1838-8528]{M.~Landriau}
\affiliation{Lawrence Berkeley National Laboratory, 1 Cyclotron Road, Berkeley, CA 94720, USA}

\author[0000-0001-7178-8868]{L.~Le~Guillou}
\affiliation{Sorbonne Universit\'{e}, CNRS/IN2P3, Laboratoire de Physique Nucl\'{e}aire et de Hautes Energies (LPNHE), FR-75005 Paris, France}

\author[0000-0002-1769-1640]{A.~de la Macorra}
\affiliation{Instituto de F\'{\i}sica, Universidad Nacional Aut\'{o}noma de M\'{e}xico,  Circuito de la Investigaci\'{o}n Cient\'{\i}fica, Ciudad Universitaria, Cd. de M\'{e}xico  C.~P.~04510,  M\'{e}xico}

\author[0000-0002-1125-7384]{A.~Meisner}
\affiliation{NSF NOIRLab, 950 N. Cherry Ave., Tucson, AZ 85719, USA}

\author{R.~Miquel}
\affiliation{Institut de F\'{i}sica d’Altes Energies (IFAE), The Barcelona Institute of Science and Technology, Edifici Cn, Campus UAB, 08193, Bellaterra (Barcelona), Spain}
\affiliation{Instituci\'{o} Catalana de Recerca i Estudis Avan\c{c}ats, Passeig de Llu\'{\i}s Companys, 23, 08010 Barcelona, Spain}

\author[0000-0002-2733-4559]{J.~Moustakas}
\affiliation{Department of Physics and Astronomy, Siena College, 515 Loudon Road, Loudonville, NY 12211, USA}

\author{A.~D.~Myers}
\affiliation{Department of Physics \& Astronomy, University  of Wyoming, 1000 E. University, Dept.~3905, Laramie, WY 82071, USA}

\author{C.~Poppett}
\affiliation{Space Sciences Laboratory, University of California, Berkeley, 7 Gauss Way, Berkeley, CA  94720, USA}
\affiliation{University of California, Berkeley, 110 Sproul Hall \#5800 Berkeley, CA 94720, USA}
\affiliation{Lawrence Berkeley National Laboratory, 1 Cyclotron Road, Berkeley, CA 94720, USA}

\author[0000-0001-7145-8674]{F.~Prada}
\affiliation{Instituto de Astrof\'{i}sica de Andaluc\'{i}a (CSIC), Glorieta de la Astronom\'{i}a, s/n, E-18008 Granada, Spain}

\author[0000-0001-6979-0125]{I.~P\'erez-R\`afols}
\affiliation{Departament de F\'isica, EEBE, Universitat Polit\`ecnica de Catalunya, c/Eduard Maristany 10, 08930 Barcelona, Spain}

\author{G.~Rossi}
\affiliation{Department of Physics and Astronomy, Sejong University, 209 Neungdong-ro, Gwangjin-gu, Seoul 05006, Republic of Korea}

\author[0000-0002-9646-8198]{E.~Sanchez}
\affiliation{CIEMAT, Avenida Complutense 40, E-28040 Madrid, Spain}

\author{M.~Schubnell}
\affiliation{Department of Physics, University of Michigan, 450 Church Street, Ann Arbor, MI 48109, USA}
\affiliation{University of Michigan, 500 S. State Street, Ann Arbor, MI 48109, USA}

\author{D.~Sprayberry}
\affiliation{NSF NOIRLab, 950 N. Cherry Ave., Tucson, AZ 85719, USA}

\author[0000-0003-1704-0781]{G.~Tarl\'{e}}
\affiliation{University of Michigan, 500 S. State Street, Ann Arbor, MI 48109, USA}

\author{B.~A.~Weaver}
\affiliation{NSF NOIRLab, 950 N. Cherry Ave., Tucson, AZ 85719, USA}

\author[0000-0002-6684-3997]{H.~Zou}
\affiliation{National Astronomical Observatories, Chinese Academy of Sciences, A20 Datun Rd., Chaoyang District, Beijing, 100012, P.R. China}

\begin{abstract}

We present spectroscopic data of strong lenses and their source galaxies using the Keck Near-Infrared Echellette Spectrometer (NIRES) and the Dark Energy Spectroscopic Instrument (DESI), \ed{providing redshifts necessary for nearly all strong-lensing applications with these systems, especially the extraction of physical parameters from lensing modeling.}  
These strong lenses were found in the DESI Legacy Imaging Surveys using Residual Neural Networks (ResNet)
and followed up by our \HST program,
\ed{with all systems displaying unambiguous lensed arcs.}
With NIRES, we target eight lensed sources at redshifts difficult to measure in the optical range and determine the source redshifts for six, between $z_s = 1.675$ and 3.332. DESI observed one of the remaining source redshifts, as well as an additional source redshift within the six systems. The two systems with non-detections by NIRES were observed for a considerably shorter 600s at high airmass.
\desied{Combining NIRES infrared spectroscopy with optical spectroscopy from our DESI Strong Lensing Secondary Target Program}, these results provide the complete lens and source redshifts for six systems, a resource for refining automated strong lens searches in future deep- and wide-field imaging surveys and addressing a range of questions in astrophysics and cosmology.


\end{abstract}

\keywords{Strong Lensing}

\section{Introduction} \label{sec:background}
\input{introduction}

\section{Observations} \label{sec:data}

\subsection{Hubble Space Telescope} \label{subsec:hst}
\input{hst}

\subsection{Keck NIRES} \label{subsec:nires}
\input{nires}


\section{Spectral Reduction} \label{sec:reduction}
\input{reduction}

\subsection{PypeIt} \label{subsec:pypeit}
\input{pypeit}

\subsection{Data Processing} \label{subsec:processing}
\input{processing}

\subsection{Redshift Fitting}

\apjed{Redshift fitting was performed by performing a Gaussian-fit to two emission lines in the spectrum. 
The two emission lines  chosen are either the strongest emission lines or the strongest from two different elements, e.g. O and H. 
The fit involved 6 parameters. The redshift $z$, which determines the mean wavelengths of the gaussians, the amplitudes $\alpha_1$, $\alpha_2$ of the gaussians, and the standard deviation of the gaussians $\sigma_1$, $\sigma_2$, as well as a constant $c$ to normalize the continuum:}

\begin{equation}
    F(\lambda) = \alpha_1 \exp\bigg[-\frac{(\lambda - \lambda_{1, \mathrm{rest}}(1 + z))^2}{2\sigma_1^2}\bigg] + \alpha_2 \exp\bigg[-\frac{(\lambda - \lambda_{2, \mathrm{rest}}(1 + z))^2}{2\sigma_2^2}\bigg] + c
\end{equation}

\apjed{where $\lambda_{1, \mathrm{rest}}$ and $\lambda_{2, \mathrm{rest}}$ are the rest-frame wavelengths of the emission lines chosen. 
This model was fit to the spectral data and uncertainties within a narrow range around the emission lines, using SciPy's \texttt{curve\_fit} routine.
The final redshift result alongside measured uncertainties are shared in \S~\ref{sec:results}.}

\section{Results}\label{sec:results}
\input{results}

\subsection{\href{https://www.legacysurvey.org/viewer/?ra=006.36439&dec=+10.1853&layer=ls-dr9&pixscale=0.262&size=101&zoom=16}{DESI J006.3643+10.1853}}\label{subsec:desi6}
\input{desi6}

\subsection{\href{https://www.legacysurvey.org/viewer/?ra=094.5639&dec=+50.3059&layer=ls-dr9&pixscale=0.262&size=101&zoom=16}{DESI J094.5639+50.3059}}\label{subsec:desi94}
\input{desi94}

\newpage
\subsection{\href{https://www.legacysurvey.org/viewer/?ra=133.3800&dec=+23.3652&layer=ls-dr9&pixscale=0.262&size=101&zoom=16}{DESI J133.3800+23.3652}}\label{subsec:desi133}

\input{desi133}

\newpage
\subsection{\href{https://www.legacysurvey.org/viewer/?ra=154.5307&dec=-00.1368&layer=ls-dr9&pixscale=0.262&size=101&zoom=16}{DESI J154.5307-00.1368}}\label{subsec:desi154}
\input{desi154}

\subsection{\href{https://www.legacysurvey.org/viewer/?ra=165.4754&dec=-06.0423&layer=ls-dr9&pixscale=0.262&size=101&zoom=16}{DESI J165.4754-06.0423}}\label{subsec:desi165}

\input{desi165}

\newpage
\subsection{\href{https://www.legacysurvey.org/viewer/?ra=215.2654&dec=+00.3719&layer=ls-dr9&pixscale=0.262&size=101&zoom=16}{DESI J215.2654+00.3719}}\label{subsec:desi215}
\input{desi215}

\section{Discussion and Conclusion}\label{sec:conclusion}
\input{conclusions}

\section*{Acknowledgment}\label{sec:acknwl}
\input{acknwl}


\bibliography{dustarchive}
\bibliographystyle{aasjournal}



\end{document}

%% file: introduction.tex

As a consequence of General Relativity, foreground galaxies or galaxy clusters can warp spacetime and bend the light of background galaxies when the alignment is nearly perfect,
resulting in the formation of multiple images, arcs, or Einstein rings \citep[e.g.,][]{walsh1979a, huchra1985a, bolton2008a}.
These strong gravitational lensing systems offer significant insights into astrophysics and cosmology. 
Since the deflection of light is a function of the lens mass, these systems may be used to study the distribution of dark matter in lensing galaxies and galaxy clusters \citep[e.g.,][]{kochanek1991a, koopmans2002a, bolton2006a, clowe2006a, koopmans2006a, bradac2008a, huang2009a, jullo2010a, shu2015a, grillo2015a, shu2016a, shu2016b, tessore2016a, shu2017a}, 
as well as dark matter substructure or light-of-sight low-mass halos \citep[e.g.,][]{vegetti2010a, hezaveh2016a, cagan-sengul2022a}. 
These systems also act as a magnifying glass for both spectral and spatial features of the lensed background galaxy, allowing more detailed study of the morphology of high redshift galaxies \citep[e.g.,][]{cornachione2018a, marshall2007a, patricio2019a, vanzella2020a}.
Lastly, multiple images of a transient in the background galaxy (typically a quasar or supernova) can be used to measure time-delays, 
providing an independent measurement of the Hubble constant $H_0$ \citep[e.g.,][]{goldstein2017a, goldstein2018a, shu2018a, wojtak2019a, pierel2019a, suyu2020a, kelly2023a, suyu2023a}.

We conducted searches for strong lensing systems \citep{huang2020a, huang2021a, storfer2024a} using deep residual neural networks (ResNet) \citep{he2015a} on the Dark Energy Spectroscopic Instrument (DESI) Legacy Imaging Surveys \citep{dey2019a}, 
and presented a large catalog of $\sim 3500$ new strong lens candidates.\footnote{The entire catalog of these lens candidates can be found on our project website \url{https://sites.google.com/usfca.edu/neuralens/}}
\desied{These candidate systems need followup observations for confirmation.
For high resolution imaging, our \HST (HST) Snapshot program 
followed up on a subset of 51 candidates. \apjed{Except for one ambiguous case, 50 of} 51 candidates in our HST sample
show unambiguous arcs, with most having multiple lensed images (Paper~I in this series, \citealt{huang2025a}).}  
Spectroscopic observations are also being carried out.
\citet{cikota2023a} published the first of our candidate systems spectroscopically confirmed by VLT MUSE.
The Dark Energy Spectroscopic Experiment \citep[DESI;][]{desi2016a, desi2016b} Strong Lensing Secondary Target Program
has already observed a large fraction of our lens candidates for spectroscopic confirmation within its 
footprint\footnote{DESI will separately follow up lensed quasar candidates discovered by autocorrelation \citep{dawes2023a}}.
\apjed{Paper~II in this series \citep{huang2025b}} will introduce this program and present strong lensing spectra from DESI Early Data Release.

\desied{For a fraction of the systems in our lens candidate catalog (preliminarily, $\sim30$\%)}, the source redshifts are too high and the typical emission features (e.g., [\ion{O}{2}] 3727) \apjed{would lie} beyond the optical range. 
We therefore turn to Keck NIRES to obtain spectroscopic redshifts for the lensed sources of these systems.
\apjed{Of these 30\% of systems, the selection criteria for the 8 systems in this paper was primarily based on visibility from Keck, the typical airmass of observation, and the brightness of the lensed arcs. Follow-up on the remaining systems are underway, and multiple Keck observing runs have been completed.}
Here, in Paper~III of this series, we present our first near-IR  spectroscopic follow-up results from Keck NIRES for systems observed by our \hst program. 
We describe our Keck NIRES observations in
\S~\ref{sec:data}, present the data reduction pipeline in \S~\ref{sec:reduction} and results in \S~\ref{sec:results},
and conclude in \S~\ref{sec:conclusion}.

 



%% file: hst.tex
For our \HST Snapshot program (GO-15867, PI: Huang),
\emph{Confirming Strong Galaxy Gravitational Lenses in the DESI Legacy Imaging Surveys}, 
we submitted a subset of our most promising lens candidates (112 targets).
Details of the program are provided in Paper I \citep{huang2025a} of this series.
Briefly, 51 systems were observed by WFC3 in F140W, with approximately half galaxy-scale and half group-scale lensing systems. 
\apjed{50 of 51 of them} were confirmed to be strong lenses based on unambiguous lensing features in the high-resolution HST images.
Our Keck NIRES observations described below target a subset of these systems.
\HST data for these systems is hosted on \href{http://dx.doi.org/10.17909/d1yz-ze43}{MAST} (DOI: 10.17909/d1yz-ze43).

%% file: nires.tex
To fully characterize a lensing system, we need to measure the redshifts of both the lens \desied{deflector} ($z_d$) and lensed source ($z_s$).
\apjed{DESI has been highly successful in obtaining lens redshifts (Paper~II, \citealt{huang2025b}), since lens spectra commonly have prominent absorption features (e.g. Ca H\&K) and the 4000 \AA{} Balmer break, within its optical range ($z_d\lesssim 1.6$). 
However, for 7 of the 8 systems in this paper, DESI was unable to obtain the source redshifts as of DESI DR2, likely since all these features, including the strong [\ion{O}{2}] doublet emission feature, are
redshifted beyond its optical range into the infrared.}
We therefore target these lensed sources using NIR spectroscopy,  namely the NIRES instrument \citep{wilson2004a} on the 10-meter Keck-2 Telescope.\footnote{We note that deeper optical spectroscopy \citep[e.g.,][]{cikota2023a} remains an alternate method for follow-up at these redshifts.}
\desied{Lensed source galaxies, being at higher redshifts, 
are typically star-forming \citep[e.g.,][]{bolton2006a}. 
NIRES provides broad wavelength coverage in the near-infrared that can be used to identify emission features of these source galaxies and secure their redshift.}
NIRES is a cross-dispersed echellette spectrograph, covering a wavelength range of 0.94~-~2.45 $\mu m$ including the Y, J, H and K-bands. 
NIRES has five spectral orders with overlap,
the only gap being between 1.85~-~1.88 $\mu m$, a region of poor atmospheric transmission. 
The slit size is 0.55 $\times$ 18$''$, 
with an average spectral resolution of $\lambda/\Delta \lambda \approx 2700$ ($\Delta v \approx 110$~km/s) 
and spectrometer pixel scale of 0.15$''$ pix$^{-1}$. 
\begin{center}
\begin{deluxetable}{c c c c c c c c c}[htbp]        
\caption{Keck NIRES Observations}
\tablehead{Date & Seeing & Airmass & Target & RA & Dec & Isophotal Magnitude & Contour Area & Exp. Time (s)}  
\startdata
\input{observing.txt}
\enddata
\tablecomments{The target naming convention is R.A. and decl. in decimal format. \desied{This naming convention differs slightly from past papers with candidates \citep{huang2020a, huang2021a} where e.g. DESI J006.3643+10.1853 is called DESI-006.3643+10.1853. We use the new convention for confirmed strong lenses, and also to match IAU naming conventions.} The isophotal magnitude (mag) is for \hst F140W in the AB system.
The area of the contour in arcsec$^{2}$ is shown as a proxy for the extent of the star forming region. *For these cases, the reported magnitudes are the lower (or brighter) bounds (see text), and only the uncertainties are shown.}
\end{deluxetable}\label{tab:observing}
\end{center}

\vspace{-7mm}
A total of eight systems from \citet[][H20]{huang2020a} and \citet[][H21]{huang2021a} that were observed and confirmed to be strong lenses by \hst 
were followed up by Keck NIRES over two half-nights.\footnote{We were unable to observe on another half-night awarded to us, on Dec 16, 2022, due to weather.} 
In this Keck NIRES program, 
our primary goal is to measure the source redshifts.
Therefore, our slits were oriented accordingly along the brightest source arc images, 
typically on high star formation rate regions (``knots") within them. 
The target name, date of observation, coordinates, exposure length, and other relevant observing information are provided in Table \ref{tab:observing}. Observations of the science targets as well as the standard star were conducted with an ABBA dither pattern with a 2~-~4$''$ nod depending on the extent of the arcs and their positions in the slit\apjed{\footnote{\apjed{The ABBA dither pattern refers to taking exposures at the first dither position (A) and then two at the second dither position (B) before returning to position A. For further information, see \href{https://www2.keck.hawaii.edu/inst/nires/night_check.html}{NIRES Instrument Documentation: Observing Procedures.}}}}. This allows a subtraction of correlated noise pattern in the spectrum by subtracting the A and B dithers. \textcolor{black}{For all systems in this paper, each dither had an exposure time of 300 seconds, and so a full ABBA set requires a 1200 second total exposure. 
The two systems DESI J023.0157-16.0040 and DESI J024.1631+00.1384 were observed at high airmass of $\approx$1.6. 
After a 600 second AB exposure each, we decided to move to the next target as we were unconvinced we could secure the redshift with the lack of visible emission features in the 2D spectrum and worsening airmass. These are the only two systems for which Keck NIRES was unable to secure a source redshift. The source redshift of DESI J023.0157-16.0040 was later captured successfully by DESI.}

We provide the magnitudes of the star-forming knots for the lensed source images of each target in Table \ref{tab:observing}. 
They are calculated for the \hst F140W band in   
the AB system, with a zeropoint of 26.45 \citep{bajaj2020a}. 
We calculate isophotal magnitudes using the total flux within a contour around a knot. 
We define the contour for each knot to be at 75\% of the peak pixel value of the knot. 
Since a dim, extended region and a bright concentrated region may have similar magnitude, we also provide the area within the contour as a proxy for the extent of the knot. 
For the two knots of DESI J023.0157-16.0040 and the fainter of the two knots covered by the slit in DESI J154.5307-00.1368, due to the presence of the lens light, 
the 75\% contours also enclose a large fraction of the lens light. 
For these cases, we instead use an elliptical aperture placed over the knot at the 75\% level as judged by the source image light profile at the far side relative to the lens location.
The magnitude estimated this way thus is an lower-bound for its true magnitude. 
More accurate photometry will be provided with full lens modeling, which will include the modeling of lens light, 
in future publications.

%% file: observing.txt
Nov 13, 2022 & 0.55$''$ & 1.38 & DESI J006.3643+10.1853 & 6.3643 & +10.1853 & 23.84 $\pm$ 0.02 & 0.0325 & 1200 \\
    & & & & & & 23.87 $\pm$ 0.02 & 0.0450 & \\
    & & 1.60 & DESI J023.0157-16.0040 & 23.0157 & $-16.0040$  & 24.02 $\pm$ 0.02* & 0.1014 & 600 \\
    & & & & & & 23.23 $\pm$ 0.02* & 0.1436 & \\
    & & 1.56 & DESI J024.1631+00.1384  & 24.1631 & +0.1384 & 22.78 $\pm$ 0.01 & 0.1925 & 600 \\
    & & & & & & 22.71 $\pm$ 0.01 & 0.2425 & \\
     & & 1.18 & DESI J094.5639+50.3059 & 94.5639 & +50.3059 & 23.85 $\pm$ 0.02 & 0.0775 & 3600 \\
     & & & & & & 24.36 $\pm$ 0.03 & 0.0325 & \\
     & & 1.43 & DESI J154.5307-00.1368 & 154.5307 &  -0.1368  & 24.24 $\pm$ 0.03* & 0.1050 & 2400 \\
     & & & & & & 25.03 $\pm$ 0.04 & 0.0350 & \\
    \hline
    Jan 10, 2023 & 0.8$''$ & 1.07 & DESI J133.3800+23.3652 & 133.3800 & +23.3652 & 23.51 $\pm$ 0.02 & 0.0350 & 2400 \\
    & & &  & & & 23.34 $\pm$ 0.02 & 0.0525 & \\
    & & 1.17 & DESI J165.4754-06.0423 & 165.4754 &  $-6.0423$  & 22.15 $\pm$ 0.01 & 0.4775 & 3600 \\
    & &  & & & & 22.28 $\pm$ 0.01 & 0.3725& \\
    & & 1.11 & DESI J215.2654+00.3719 & 215.2654 & +0.3719 & 24.15 $\pm$ 0.03 & 0.0925 & 2400 \\
    & & & & & & 24.41 $\pm$ 0.03 & 0.0800 & 

%% file: reduction.tex
We use PypeIt for spectral reduction.
We introduce this open-source package in \S~\ref{subsec:pypeit} and our reduction process in \S~\ref{subsec:processing}.

%% file: pypeit.tex
PypeIt is a versatile Python package for semi-automated reduction of spectroscopic data \citep{prochaska2020a, prochaska2020b}. 
The reduction takes several steps. 
First it performs overscan subtraction, dark-current and bias correction, bad pixel masking, slit edge detection, and wavelength calibration on both the science and standard star observations. 
Finally, PypeIt performs Poisson-limited sky subtraction and object extraction to generate the 1D and 2D science spectra. 
Object detection is typically done automatically, but for faint galaxies, where redshift measurements  depend on emission line detection, 
a manually laid trace is used to extract the spectra. These science spectra can be flexure corrected, coadded, and even telluric-corrected. For further information regarding PypeIt, we direct the reader to \citet{prochaska2020a, prochaska2020b}. 

%% file: processing.tex
For this paper, the spectra taken for each night consisted of science observations of the lensed sources and standard star observations from the CALSPEC database \citep{bohlin2014b}. 
Flat frames (120~sec $\times$ 10) and dark frames (120~sec $\times$ 10) were taken for calibration.
Wavelength calibration was performed using sky lines.\footnote{\url{https://pypeit.readthedocs.io/en/release/calibrations/wave_calib.html}} 
PypeIt v1.13.0 was used for this analysis. 
\desied{Applying PypeIt} to a broad range of spectra requires appropriate manual adjustments for our specific use-case: spectroscopic extraction of source galaxy emission lines.

The main differences between our procedure and a standard PypeIt reduction are the use of a manual object extraction and the omission of a telluric correction. 
The continuum of the lensed source arcs typically does not have sufficient signal to be separated from noise.  
This means the emission features are not easily detected by PypeIt using the automatic object detection, and requires the manual specification of the object position and an estimation of the FWHM.
Additionally, the automatic extraction routine had to be suppressed for a particular system (See \ref{subsec:desi133}). This suppression was later adopted for all systems.
Telluric correction is not used due to the low signal-to-noise ratio (SNR) for our spectra (J.\ Hennawi, priv.\ comm.). 
At low SNR, this correction exhibits pathological behaviors as it is challenging to distinguish noise from signal. 
This is acceptable, since for our targets, the telluric-correction does not impact the redshift determination based on identifying emission features.

The standard star used to generate the sensitivity function for both nights was of the G191-B2B observations from the night of Jan 10, 2023.\footnote{Although standard star observations of BD+75325 were taken on the night of Nov 13, 2022, PypeIt faced an internal error in the generation of the sensitivity file using this standard star, and the issue  was raised to the PypeIt team (J.\ Hennawi, X.\ Prochaska, priv.\ comm.)} 
These were taken with an exposure time of 15 sec each in an ABBA dither pattern of 2$''$. 
Using the estimate of the sensitivity function from a different night may slightly affect line ratios, but should not impact redshift measurements. The ABBA frames of the emission line spectra are individually extracted and coadded with their positive and negative features using PypeIt's weighted 1D coaddition pipeline \citep{prochaska2020b}. 
\desied{PypeIt internally performs both correction of the spectra to the barycentric frame as well as to their vacuum wavelengths.}

For all systems, we place the NIRES slit orientation along a lensed arc of the source galaxy in order to capture in the same slit
two lensed images of the same star formation region 
or two different star formation regions (\S~\ref{subsec:desi6} \& \ref{subsec:desi154}) of the source galaxy.
Therefore the detected emission features have identical 
(or very similar, \S~\ref{subsec:desi154})
redshifts, allowing us to coadd their individual emission spectra. 
The combined 1D spectra 
are inspected alongside the 2D spectra (see, e.g., Figure \ref{fig:desi6}). 

%% file: results.tex
We present the six lensing systems for which we have successfully obtained the source redshifts from Keck NIRES spectra, 
ranging from \apjed{1.67511 to 3.33185} in \S~\ref{subsec:desi6}~-~ \ref{subsec:desi215}.
The results are summarized in Table~\ref{tab:results}, alongside DESI spectra of the lens galaxies and other source galaxies with emission features within the DESI optical range, for completeness. 
\apjed{The uncertainties from redshift fitting for each system, of order O($10^{-4}$) to O($10^{-5}$), are also provided.
Since these uncertainties are small, we also note possible systematic uncertainties from PypeIt wavelength calibration.
We find an RMS uncertainty of $ < 0.2$ pixels for all systems tested. This corresponds to a maximum uncertainty of 0.6\AA, which would cause a maximum of O($10^{-4}$) systematic error in redshift.}
\apjed{For DESI redshifts, as} described in \citealt{desi2024a} (D24), random uncertainties vary. For DESI BGS's ($z < 0.4$, with $z_\textrm{med} \sim 0.2$) uncertainties are $\sigma_z =
0.00003(1 + z)$ (D24, Table 6). For LRG's ($0.4 < z < 1.1$), the uncertainties are estimated to be $\sigma_z = 0.00014(1 + z)$ (D24, Table 5). Lastly, for  ELG's ($0.6 < z < 1.6$), such as the two source galaxies observed by DESI in Table~\ref{tab:results}, we have $\sigma_z = 0.000026(1 + z)$ (D24, Table 5).

\begin{center}
\begin{deluxetable}{cccc}[htbp]                
\caption{Redshift Results}
\tablehead{System Name &  $z_d$ & $z_s$ & \apjed{$\sigma_{z_s}$}}  
\startdata
\input{zs-results.txt}
\enddata 
\tablecomments{
We report redshifts $z_s$ of two separate sources for DESI J006.3643+10.1853 (See \S~\ref{subsec:desi6}). We report two redshifts $z_s$ for two star-forming knots within the same source for DESI J154.5307-00.1368 (See \S~\ref{subsec:desi154}). \ed{$^a$Redshifts determined by DESI. $^b$Redshift determined from Lick. Rest are from Keck NIRES \textbf{(this work)}.}}
\end{deluxetable}\label{tab:results}
\end{center}


\vspace{-12mm}
The
spectroscopic redshifts for the lenses from the DESI Strong Lens Secondary Target program \apjed{(described further in Paper II, \citealt{huang2025b})} is shown for all systems except for DESI J094.5639+50.3059 and DESI J023.0157-16.0040.
For the former, we show our redshift result from the Kast spectrograph on the 3-m Shane Telescope at the Lick Observatory on Feb 6, 2022, 
for which \citet{tran2022a} also reported the same value.
Given that these are elliptical galaxies, these spectroscopic redshifts agree well with photometric redshifts from \citet{zhou2020a}. 
The two systems (DESI J023.0157-16.0040 and DESI J024.1631+00.1384) for which we have not secured the source redshifts with Keck NIRES only had 600 sec of exposure at higher airmass (about 1.6), shorter than the other systems (1200 - 3600 sec), at airmasses of about 1.1-1.4.
\ed{DESI J023.0157-16.0040 was later targeted successfully by DESI at the limits of its optical range, and the source redshift was determined to be $1.5818$. 
DESI J024.1631+00.1384 will be targeted again in future observations.}


%% file: zs-results.txt
    DESI J006.3643+10.1853 & 0.4631$^a$ & \textbf{2.39688} & \apjed{$\pm$ 0.00004}\\
     &  & 1.3143$^a$ & \apjed{$\pm$ 0.00006$^a$}\\
 DESI J023.0157-16.0040 & - & 1.5818$^a$ & \apjed{$\pm$ 0.00007$^a$}\\
DESI J024.1631+00.1384 &   0.3445$^a$ & - & -\\
     DESI J094.5639+50.3059 &  0.522$^b$ & \textbf{3.33185}& \apjed{$\pm$ 0.00010} \\
    DESI J133.3800+23.3652 &  0.3053$^a$  & \textbf{2.18858} & \apjed{$\pm$ 0.00002}\\
DESI J154.5307-00.1368 &  0.3718$^a$  & \textbf{1.73735, 1.73885}  & \apjed{$\pm$ 0.00004, 0.00006}\\
    DESI J165.4754-06.0423 &  0.4834$^a$  & \textbf{1.67511} & \apjed{$\pm$ 0.00005}\\
    DESI J215.2654+00.3719 &  0.6566$^a$  & \textbf{2.20645} & \apjed{$\pm$ 0.00035}

%% file: desi6.tex
\begin{figure}[!ht]
  \centering
  \includegraphics[width=\textwidth]
  {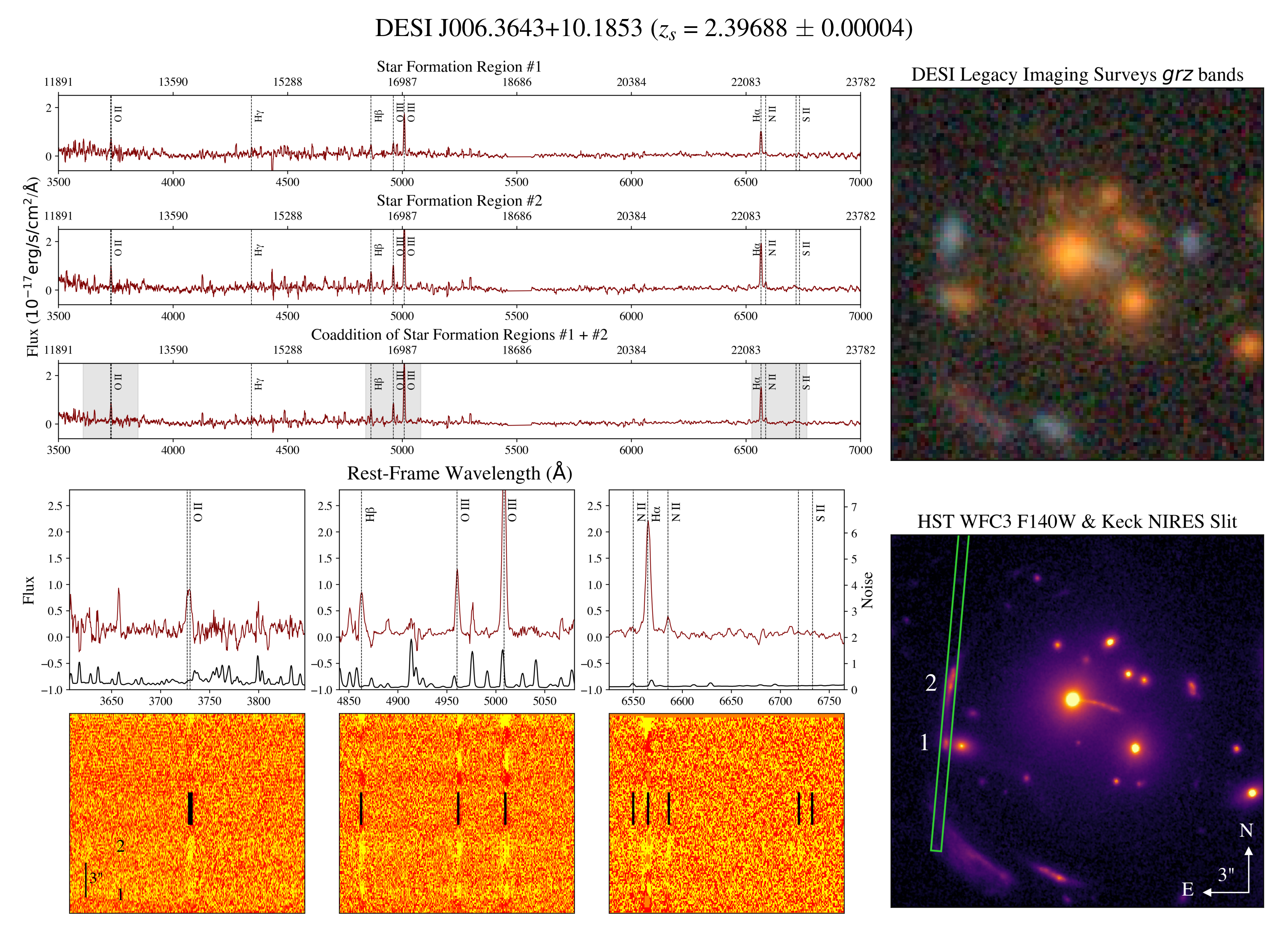}
  \caption{DESI J006.3643+10.1853. \textbf{Left Column}: The first and second rows show the spectra for each source image \pound1 and \pound2. 
  The third row shows the weighted coaddition of the source image pair. 
  The spectra in these panels are smoothed with a kernel of 10~pixels.
  The fourth row shows the zoom-in for individual emission lines (maroon), 
  alongside the error spectrum (black), 
  with a convolution kernel of 5~pixels. \ed{The error spectrum is provided on a separate scale for clarity.}
  We show the 2D spectra in the bottom row, with black lines showing the alignment with the zoomed-in emission lines on the fourth row above. 
  The locations of the source arcs are indicated (``1" and ``2" in black) in the 2D spectra.
  The spatial separation of the star forming regions in 2D spectra (note the scale bar corresponding to 3$''$) matches the separation in the \hst F140W image.
  Finally, for the negative signals of the spectral features, the bright pixels above and below to them (especially prominent for [\ion{O}{3}] doublet) are due to an artifact of the PypeIt 2D coaddition pipeline. 
  \textbf{Right Column:} We show the discovery image from the DESI DR9 Legacy Surveys Imaging ($grz$ bands) and the \hst F140W image with the slit orientation (green). 
  The source arcs are labeled as ``1'' and ``2'' in white.
  The length of the compass corresponds to 3$''$.}
  \label{fig:desi6}
\end{figure}

This system was discovered in H21 and the Legacy Imaging Surveys image was given a human inspection score of 2.5.
In H21, The lens candidate systems with a human scores of 2 or 2.5 were given a C~grade,
but they stated that all systems with a human score of 2.5 is at least a likely lensing system. 
In the DESI Legacy Surveys image (Figure~\ref{fig:desi6}), 
there are four arclets with very similar blue colors,
even though they do not form a ``classic'' quad pattern.
But this is not a surprise, given the lens appears to be at least a group or a small cluster. 
The \hst image confirms that three of the four suspected blue lensed images indeed have the appearance of lensed arcs, curving toward the brightest galaxy in the group, 
each arc with two star formation ``knots''.
The fourth arc appears to be a rare prominent radial arc \citep[see e.g.,][]{broadhurst2000a} near and pointing toward the brightest galaxy in the group.
Keck NIRES spectroscopy now provides the redshift of the brightest lensed image. 
In Figure \ref{fig:desi6}, we find clear evidence of the emission features H$\alpha$, [O II], [O III], H$\beta$ and [N II] in both the 1D and 2D spectra.
The source redshift is determined to be \apjed{$z_s = 2.39688$}.
The 1D spectra of the two knots are both shown in Figure~\ref{fig:desi6} (which are at the same redshift as expected)
as well as their coaddition.
This lensed source is among the brightest sources targeted in this Keck program. 
Besides the emission features, faint traces corresponding to the continuum of the source spectra can be seen in the 2D spectra (Figure~\ref{fig:desi6}).
\citet{tran2022a} reported the same source redshift of 2.396 for this system, though they did not show the spectra.
\ed{There also appears to be a faint reddish arc to the SE of the Legacy Surveys image. This reddish arc is confirmed by \hst, and has DESI spectra with a strong [\ion{O}{2}] feature.
The redshift of this second lensed source is at $z = 1.3143$.}

%% file: desi94.tex
\begin{figure}[!ht]
  \centering
  \includegraphics[width=\textwidth]{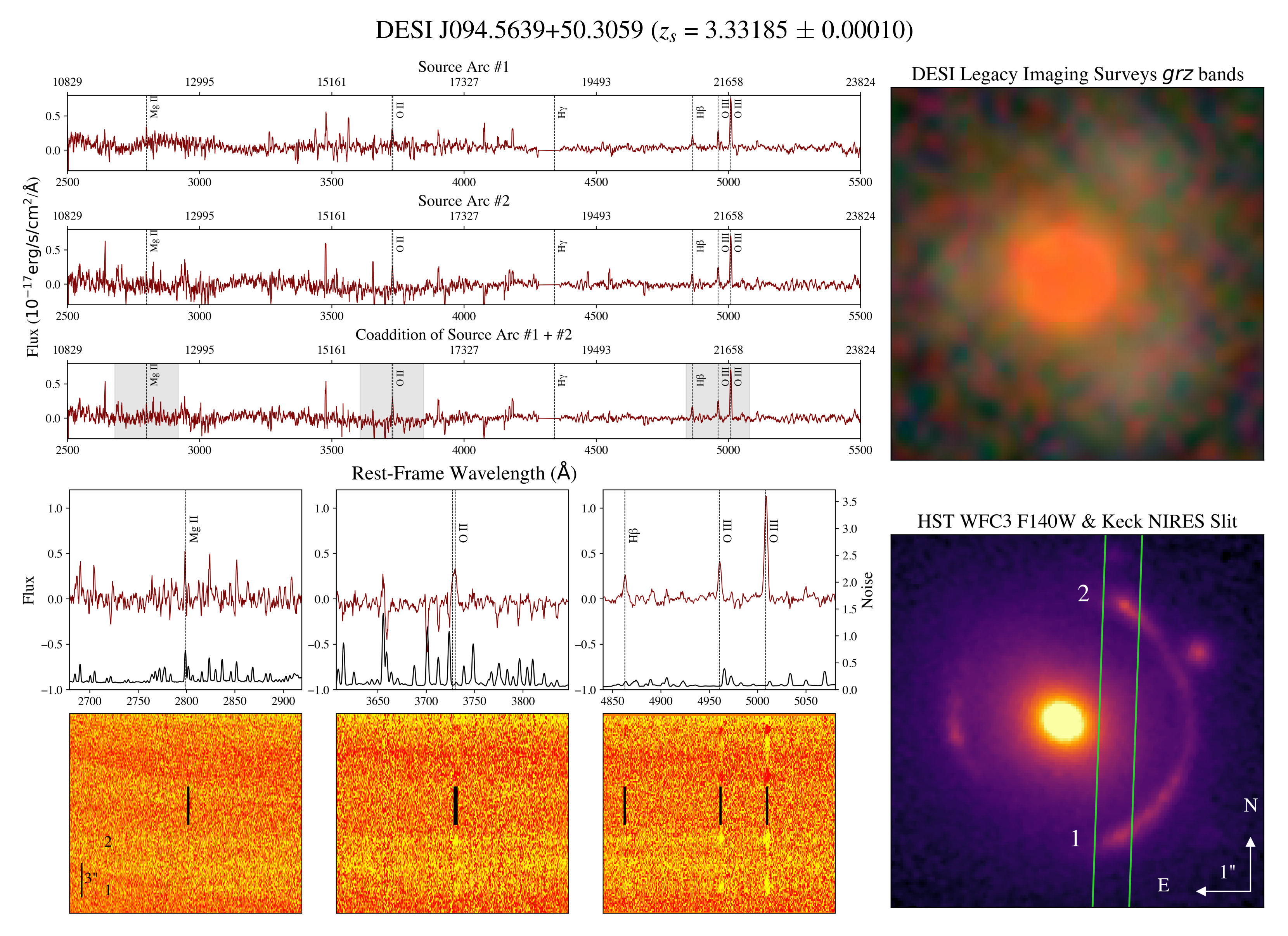}
  \caption{DESI J094.5639+50.3059. 
  For the arrangement of the panels, see the caption of Figure~\ref{fig:desi6}.}
  \label{fig:desi94}
\end{figure}

DESI J094.5639+50.3059 was given an A Grade in H21 owing to the clear presence of the main arc to the west of the foreground galaxy 
and a counterarc on the opposite side (Figure~\ref{fig:desi94}).
We observed this system using the Kast spectrograph on the 3-meter Shane Telescope at the Lick Observatory on Feb 6, 2022.
We determined the lens redshift to be $z_d = 0.522$ based on clear Ca~H\&K, G, and Mg absorption lines.
But we were unable to obtain the source redshift using Kast at Lick.
The lens redshift reported by \citet{tran2022a} is in agreement with this.
For the lensed source, from our Keck NIRES spectra, we find clear evidence of [\ion{O}{2}], [\ion{O}{3}], 
H$\beta$ emissions in the 1D and 2D spectra, not just for the coaddition but also for the individual images of the knot at the tips of the main arc. 
There could also be a detection of a \ion{Mg}{2} emission line, though it aligns with the noise (in black) and is not visible in the 2D spectrum.
It has the highest source redshift in this paper, at \apjed{$z_s = 3.33185$. This system has been modeled in Paper V (see \citet{huang2025c}).}

%% file: desi133.tex
DESI J133.3800+23.3652 was given Grade B in H21.  
The spectrum of the source arcs show  evidence of strong H$\alpha$, H$\beta$, [\ion{O}{2}] and [\ion{O}{3}] doublet features in the coadded 1D and 2D spectra.
The high SNR in this system allows us to infer the presence of the much weaker [\ion{N}{2}] and [\ion{S}{2}] features as well. 
We determine the source redshift of this system to be \apjed{$z_s = 2.18858$}. 
The two source images on opposite sides of the foreground galaxy have the same redshifts, 
providing further support for the lensing nature of this system.
DESI J133.3800+23.3652 reduction faced a unique PypeIt error in which a portion of the 1D spectrum from the automatically detected lens galaxy was copied onto the trace corresponding to the second source image. 
In order to suppress this unusual behavior, 
PypeIt automatic object detection was turned off.\footnote{This issue was also raised to and acknowledged by the PypeIt team (X.\ Prochaska, priv.\ comm.)
} 
This enabled us to perform a manual extraction on the source galaxy spectra for both images of the star forming knot using PypeIt's manual extraction feature.
The manual extraction routine was then adopted for all other systems as well, \desied{as mentioned in} 
\S~\ref{subsec:processing}. There is a second lensed source to the East of the measured lensed source which falls into the slit. However, we were unable to locate any emission features for that source within our 1200s exposure. 

\begin{figure}[ht]
  \centering
  \includegraphics[width=\textwidth]{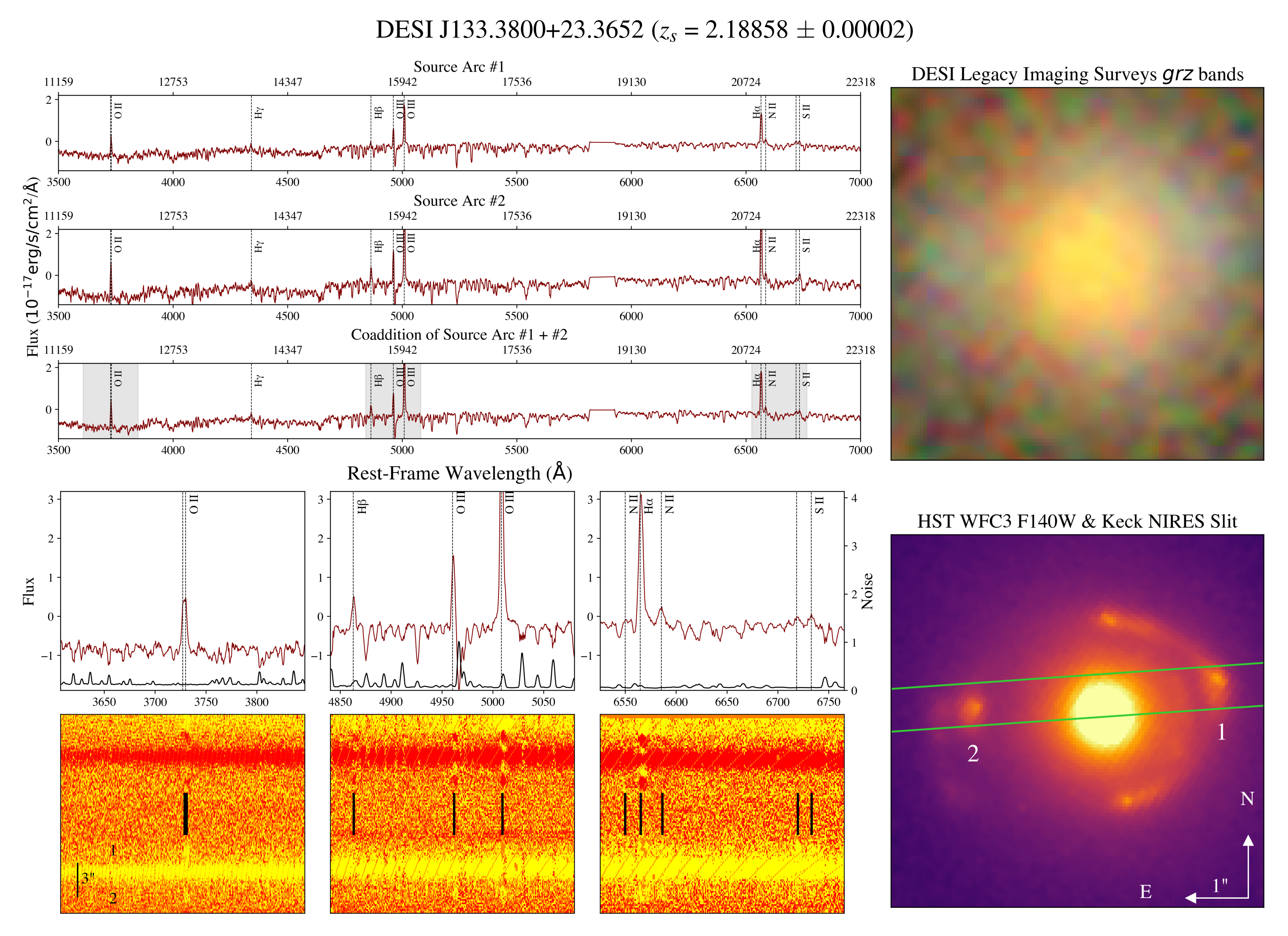}
  \caption{DESI J133.3800+23.3652. 
  For the arrangement of the panels, see the caption of Figure~\ref{fig:desi6}.}
  \label{fig:desi133}
\end{figure}

%% file: desi154.tex
\begin{figure}[ht]
  \centering
  \includegraphics[width=\textwidth]{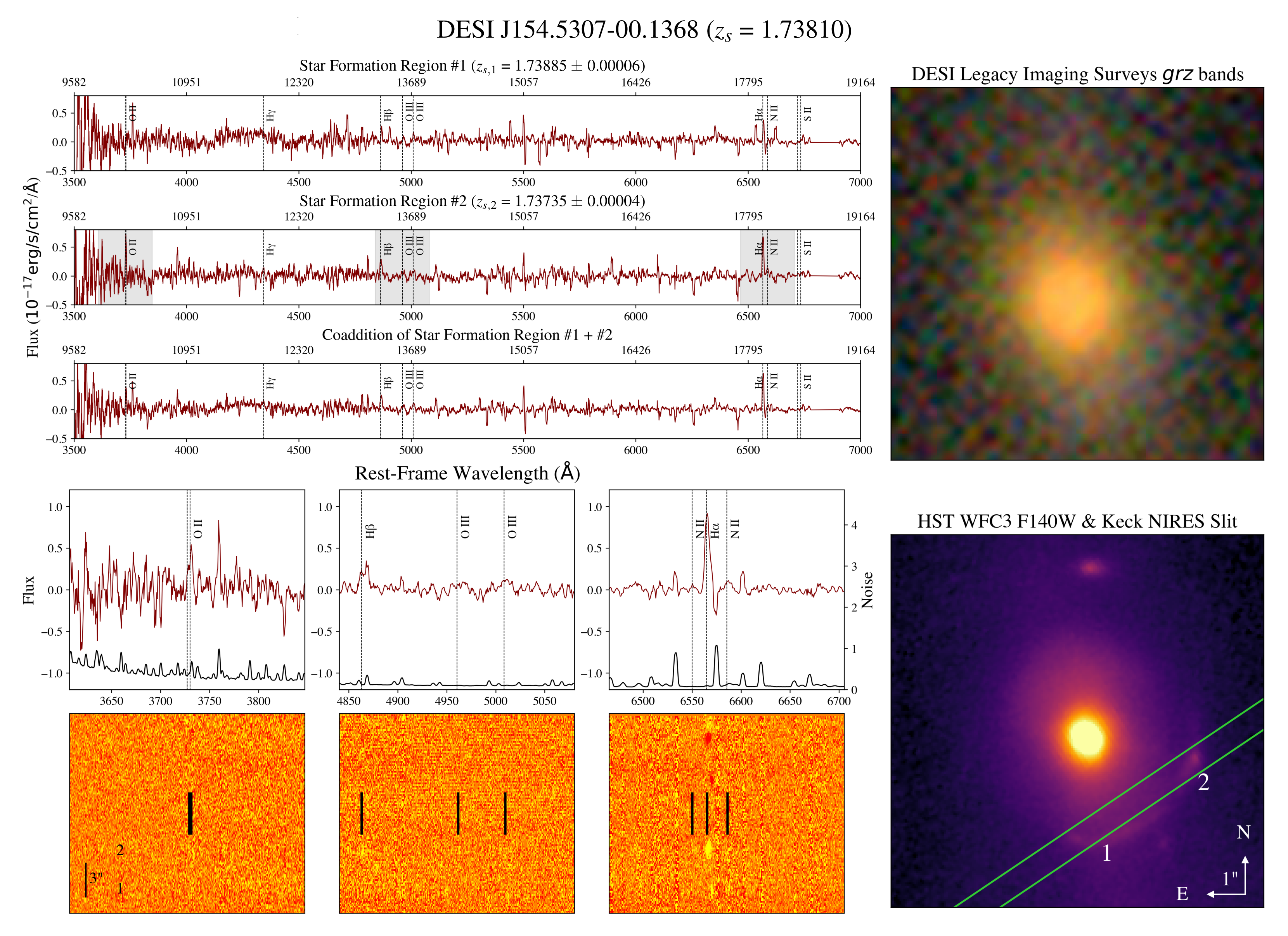}
  \caption{DESI J154.5307-00.1368. For the arrangement of the panels, see the caption of Figure~\ref{fig:desi6}. 
  Note that in the 2D zoom-in, the H$\alpha$ emission features corresponding to the two star forming knots do not align exactly in the spectral direction, 
  indicating slightly different redshifts for \apjed{knot \pound1 and \pound2 at $z_{s, 1} = 1.73885$ and $z_{s, 2} = 1.73735$, respectively}. 
  For this system, the zoom-in panels in the fourth row highlight the 1D spectrum of star formation region \pound2 (corresponding to the gray highlighted areas in the second row) and not the coaddition.}
  \label{fig:desi154}
\end{figure}

This system was given Grade B in H20, 
on account of the faint blue arc to the SW of the foreground galaxy (Figure~\ref{fig:desi154}). 
The NIRES 2D spectra show a strong signal for H$\alpha$ emission, 
and next to it, a detection of the [\ion{N}{2}] emission line.
The presence of H$\beta$ is weaker but clear, but we do not find an obvious detection of the [\ion{O}{3}] doublet.
\textcolor{black}{Finally, there may be a hint of the [\ion{O}{2}] doublet in the 2D spectrum, which can also been seen in one of the coadded 1D spectra (second row in Figure~\ref{fig:desi154}), at $\sim2\sigma$. 
We note that this feature is only found in one of the two star forming regions, which may be due to low signal-to-noise ratio and/or 
the two knots having different oxygen compositions.}
The redshift of the system is determined to be \apjed{$z_s = 1.73810$}.
Note that the two tips of the arc are two \emph{different} star formation regions. 
They have slightly different redshifts, at \apjed{1.73885} and \apjed{1.73735}, 
for the eastern and western knot respectively (labeled as knot \pound1 and \pound2 in Figure~\ref{fig:desi154}).
\ed{We see clear evidence for this difference in the H$\alpha$ emission feature in both 2D and 1D spectra.
The H$\beta$ emission appears to show the same difference.}
\apjed{This difference in redshift corresponds to a line-of-sight velocity difference of 164.2 km/s \cite[Eq. 10,][]{hogg99}. 
The existence of a velocity difference indicates these may be two separate knots, rather than multiple images of the same knot.
Lens modeling of this system is forthcoming, and may shed more light on the specific lensing configuration. 
For now, we expect that the system is either a singly lensed arc (the targeted south-west arc) or an ``inclined double configuration" \citep{saha2003}, where the second arc is the source north of the lens galaxy.
Depending on the outcome, one interpretation may be that separate, lensed knots are on opposite ends of the galaxy, and we detect the velocity difference due to galactic rotation.}
We average the redshifts of the two knots to determine the source redshift.
\textcolor{black}{Approximately 5$''$ to the north there appears to be an arclet that may be another lensed image of the same background galaxy. 
We will target this object in future spectroscopic observations.}

%% file: desi165.tex
\begin{figure}[ht]
  \centering
  \includegraphics[width=\textwidth]{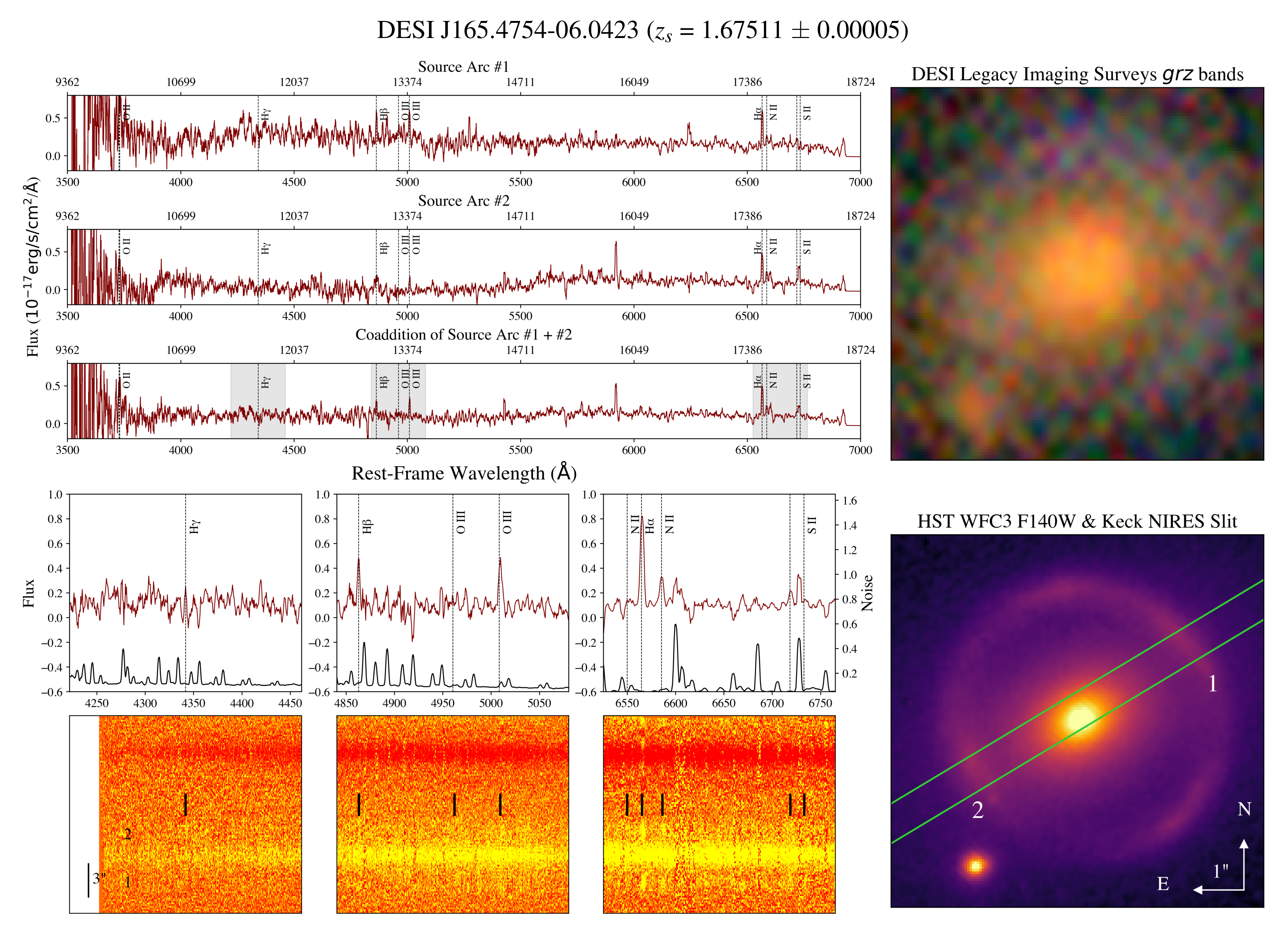}
  \caption{DESI J165.4754-06.0423. 
  For the arrangement of the panels, see the caption of Figure~\ref{fig:desi6}.}
  \label{fig:desi165}
\end{figure}

DESI J165.4754-06.0423 was given Grade C in H21, 
with a human score of 2.5,
which indicates at least a likely candidate (see \S~\ref{subsec:desi6}).
We determine the source redshift to be \apjed{$z_s = 1.67511$}.
It is the lowest redshift source galaxy in this work,
with the [\ion{O}{2}] doublet emission feature
lying just beyond the optical edge of DESI spectroscopy. 
From the NIRES spectra, clear H$\alpha$, [\ion{N}{2}], and H$\beta$ signals, 
alongside the brighter of the [\ion{O}{3}] doublet and a possible weak [\ion{S}{2}] and H$\gamma$ signal enable us to secure the redshift of the source galaxy. 
The bump in the 1D coadded spectrum between the [\ion{S}{2}] doublet is due to imperfect subtraction of a bright sky line.
This can be seen as a jump at that wavelength 
in the noise spectrum, which is only 
present in one of the two individual 1D spectra (Source Arc \pound2 in Figure~\ref{fig:desi165}).

%% file: desi215.tex
\begin{figure}[hb]
  \centering
  \includegraphics[width=\textwidth]{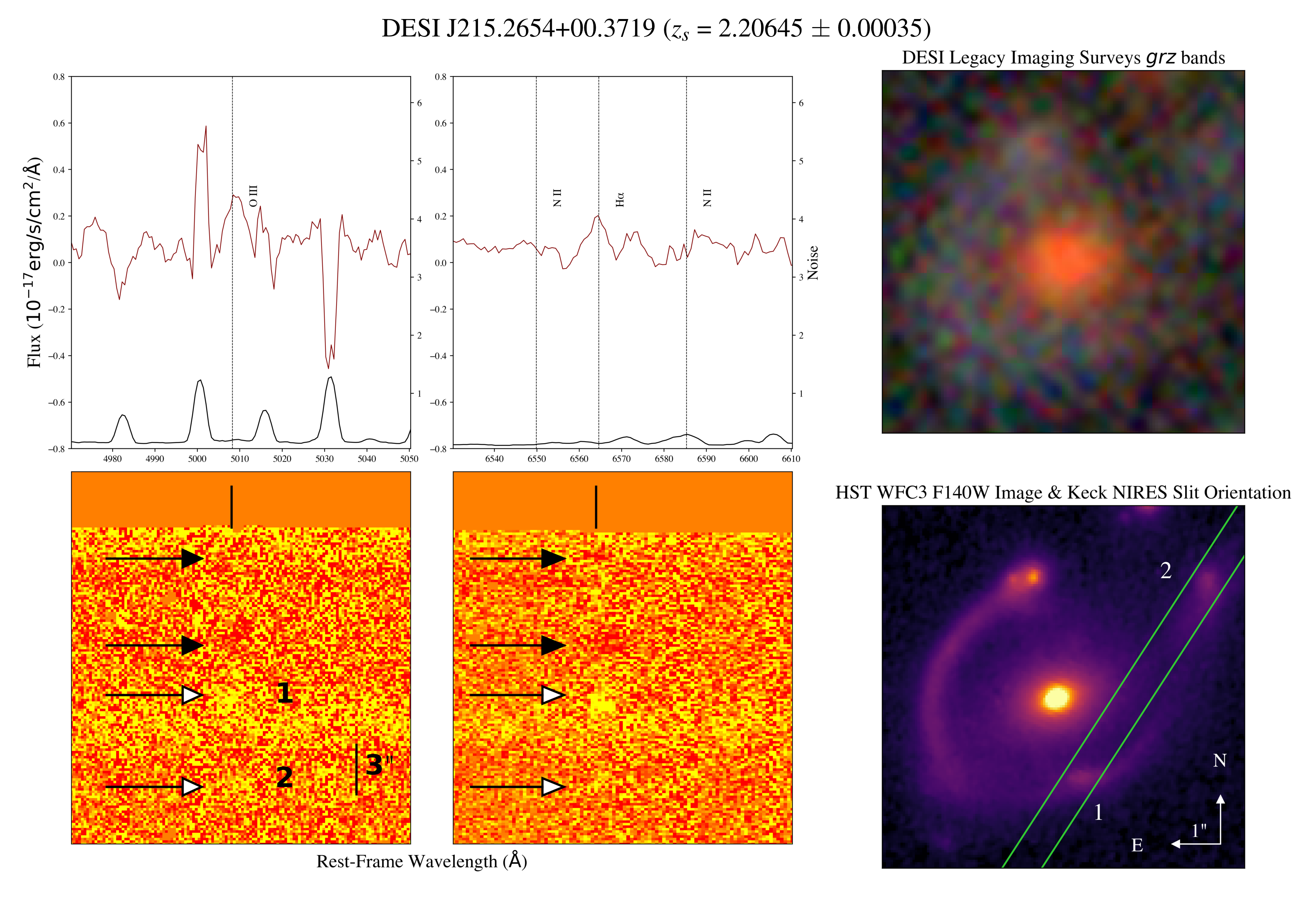}
  \caption{DESI J215.2654+00.3719. 
  \textbf{Left Column}: 
   In the top row, the left and right panels show the zoom-in for the coadded 1D emission lines [\ion{O}{3}] and H$\alpha$ (maroon), respectively, 
   alongside the error spectrum (black), with a convolution kernel of width 2 pixels. 
  The 2D spectra are presented in the bottom row, 
  with vertical lines near the top showing the alignment with 1D emission lines. 
  The locations of the source arcs are indicated (``1" and ``2" in black) in the lower left 2D spectrum.
  In the 2D spectra, we highlight the positive (white arrowheads) and negative signals (black arrowheads) from the dither pattern.
  The presence of both distinguishes the pattern from noise.
  The spatial separation of the star forming regions in 2D spectra matches that in the \hst F140W image (4.86$''$).
  Also note the presence of the continuum of the lensing galaxy spectrum between the emission features \pound1 and \pound2. 
  \textbf{Right Column:} We show the discovery image from the DESI DR9 Legacy Surveys Imaging ($grz$ bands) and the \hst F140W image with the slit orientation (green). 
  The source arcs are labeled as ``1'' and ``2'' in white.}
  \label{fig:desi215}
\end{figure}

DESI J215.2654+00.3719 is a Grade A candidate in H20,
with two prominent blue arcs around an elliptical galaxy.
This system has the faintest star forming regions in this program, 
with the brightest knot in the arcs being 22.8~mag in \hst F140W.\footnote{We did not target the brightest feature in the F140W cutout image near the top center (Figure~\ref{fig:desi215}, low right panel).
Given its reddish color from the Legacy Surveys $grz$ image, 
we suspect this feature likely corresponds to a small interloper or satellite galaxy. 
We will address the nature of this object in future followup observations.}
It was observed  with 2400~sec of exposure because it was near the end of night.
With careful inspection, we find two weak features in the 2D spectrum, 
with both a visible positive and negative signal to ensure they are not due to noise 
(Figure~\ref{fig:desi215}). 
We reason that they must be the strongest features in the NIR range, 
and test to fit them for H$\alpha$, [\ion{O}{3}], and [\ion{O}{2}] in order. 
We find that the spacing between them exactly matches that between H$\alpha$ and the brighter of the [\ion{O}{3}] doublet, 
which are generally the two 
emission features with the clearest detections
in the other NIR spectra presented in this work. 
This places the source galaxy at a redshift of \apjed{$z_s = 2.20645$}. 
\apjed{The non-detection of other emission features such as [\ion{O}{2}], H$\beta$, [\ion{N}{2}] and [\ion{S}{2}] is unsurprising due to the intrinsic relative faintness of these emission lines compared to the faint detected H$\alpha$ and [\ion{O}{3}] features and given the intrinsically low brightness star-forming regions (magnitude of 24.15 \& 24.41, see Table \ref{tab:observing}).}
The zoom-in views of the 1D spectrum and 2D features are shown in Figure \ref{fig:desi215}.

%% file: conclusions.tex
\subsection{Further Analysis}
\apjed{The focus of  this paper is to obtain redshifts, which are necessary for setting the overall scale for the lens mass profiles through lens modeling \citep[Paper V;][]{huang2025c}. 
However, the spectra contain significant information beyond redshift determination. 
Other possible further research include analysis of high-redshift galaxy star formation rates through measurement of equivalent widths of lines such as H$\alpha$ and H$\beta$ \citep[E.g.][]{livermore2015, patricio2018, nagy2023}. 
Given that the continuum is not always detected, this would represent a lower bound, which nevertheless is a useful estimate. 
This information is also essential for estimating supernova rates, and can be used to identify ideal systems for lensed-supernova searches \citep[E.g.][]{shu2018a, shu2021a}. 
Lastly, given that the strong lensing systems in the paper magnify high-redshift source galaxies at redshifts $z \sim 1.7-3.3$, their metallicities and galaxy evolution \citep[E.g.][]{stark2013a, yuan2013, patricio2019, he2024} could be studied, as well as reionization in general, through further Ly$\alpha$ studies \citep{rivera2017, khullar2021}.}

\subsection{Conclusion}
We  present and analyze Keck NIRES spectra of strong lensing systems discovered in the DESI Legacy Imaging Surveys using residual neural networks.
\textcolor{black}{Redshifts from this program can be used for various strong lensing applications, such as characterizing the selection function of the neural network used, and converting geometrical parameters to physical ones.
Additionally, this paper demonstrates the potential of the neural network lens sample for discoveries of high-redshift galaxies, galaxy formation studies, and understanding reionization.} The candidate systems used have unambiguous lensing features that have been confirmed by our \HST program (GO-15867, PI: Huang). 

We targeted the lensed sources of these systems because they were beyond the optical range of DESI, specifically when the [\ion{O}{2}] doublet at 3727 \ang is redshifted beyond 9800 \ang \citep{desi2023a}, at $z \sim 1.6$. The exception is DESI J023.0157-16.0040, where the [\ion{O}{2}] doublet is barely within range and was successfully observed by DESI to have a redshift of 1.5818.
We have succeeded in measuring the lensed source redshifts for six (with an exposure time of 1200 - 3600 sec) of seven other observed systems,
\apjed{with $z_s$ between = 1.67511 and 3.33185.}
Two each of these six systems were given Grade A, B, and C 
(with a human inspection score of 4, 3, and 2.5, respectively)
in our lens searches \citep[][where we pointed out that candidates with a score of 2.5 or above are at least likely lenses]{huang2020a, huang2021a}. \ed{The two systems for which we did not obtain the source redshifts with Keck NIRES
had considerably shorter exposure times (600 sec) and were at higher airmass ($\sim$1.6). DESI J024.1631+00.1384 will be targeted in future observations.}
This is a relatively small sample
and we will present larger samples of confirmed systems in the \desied{DESI Strong Lens Foundry} series.
Nevertheless this is a promising first assessment that attests to the quality of our lens candidates found in the DESI Legacy Imaging Surveys. These NIR spectroscopic observations show great promise for obtaining lensing systems at high redshifts, for which the strongest spectral features of the lensed sources are beyond the optical range.

Though the main objective of this program is to obtain redshifts,
the spectra we obtained clearly have sufficient signal-to-noise for follow-up investigations, 
in particular, 
measuring the equivalent width of the most prominent emission lines \apjed{and determining metallicity ratios. 
This would allow us to estimate the star formation rates and therefore supernova rates in the lensed source galaxies, and study the metallicity and formation process of high-redshift galaxies. This work is beyond the scope of this paper, and may be carried out in a future study.}

For systems observed with at least one full ABBA dither pattern (1200~sec), Keck NIRES achieves a 100\% success rate in obtaining redshifts for the lensed sources.
In addition to NIR spectroscopy, 
another possible method for follow-up is to use deep optical spectroscopy to identify weaker spectral lines \citep[e.g., DESI J253.2534+26.8843,][]{cikota2023a}.
In future follow-up observations, 
the comparison between the efficiencies of  NIR spectroscopy and deep optical spectroscopy for high redshift lensed sources could be \desied{investigated further with our large sample of lens candidates}.

%% file: acknwl.tex
This work was supported in part by the Director, Office of
Science, Office of High Energy Physics of the US Department
of Energy under contract No.\ DE-AC025CH11231. 
This
research used resources of the National Energy Research
Scientific Computing Center (NERSC), a U.S. Department of
Energy Office of Science User Facility operated under the same
contract as above and the Computational HEP program in The
Department of Energy's Science Office of High Energy Physics
provided resources through the ``Cosmology Data Repository''
project (grant No.\ KA2401022).
S.A. acknowledges the opportunities provided by the Undergraduate Research Apprenticeship Program (URAP) at the University of California, Berkeley.
X.H.\ acknowledges the University of San Francisco Faculty Development Fund. 
Support for HST program 15867 was provided by NASA through a grant from the Space Telescope Science Institute, which is operated by the Association of Universities for Research in Astronomy, Inc., under NASA contract NAS 5-26555.
We thank the creators of PypeIt for support with usage of the package.
We also thank the Support Astronomers and the technical staff of the W.M\ Keck Observatory. 
We recognize the significant cultural role of Mauna Kea within the indigenous Hawaiian community, and we appreciate the opportunity to conduct observations from this revered site.

The Legacy Surveys consist of three individual and complementary projects: the Dark Energy Camera Legacy Survey (DECaLS; Proposal ID \#2014B-0404; PIs: David Schlegel and Arjun Dey), the Beijing-Arizona Sky Survey (BASS; NOAO Prop. ID \#2015A-0801; PIs: Zhou Xu and Xiaohui Fan), and the Mayall z-band Legacy Survey (MzLS; Prop. ID \#2016A-0453; PI: Arjun Dey). DECaLS, BASS and MzLS together include data obtained, respectively, at the Blanco telescope, Cerro Tololo Inter-American Observatory, NSF’s NOIRLab; the Bok telescope, Steward Observatory, University of Arizona; and the Mayall telescope, Kitt Peak National Observatory, NOIRLab. Pipeline processing and analyses of the data were supported by NOIRLab and the Lawrence Berkeley National Laboratory (LBNL). The Legacy Surveys project is honored to be permitted to conduct astronomical research on Iolkam Du’ag (Kitt Peak), a mountain with particular significance to the Tohono O’odham Nation.
NOIRLab is operated by the Association of Universities for Research in Astronomy (AURA) under a cooperative agreement with the National Science Foundation. LBNL is managed by the Regents of the University of California under contract to the U.S. Department of Energy.
This project used data obtained with the Dark Energy Camera (DECam), which was constructed by the Dark Energy Survey (DES) collaboration. Funding for the DES Projects has been provided by the U.S. Department of Energy, the U.S. National Science Foundation, the Ministry of Science and Education of Spain, the Science and Technology Facilities Council of the United Kingdom, the Higher Education Funding Council for England, the National Center for Supercomputing Applications at the University of Illinois at Urbana-Champaign, the Kavli Institute of Cosmological Physics at the University of Chicago, Center for Cosmology and Astro-Particle Physics at the Ohio State University, the Mitchell Institute for Fundamental Physics and Astronomy at Texas A\&M University, Financiadora de Estudos e Projetos, Funda\c{c}\~ao Carlos Chagas Filho de Amparo \`a Pesquisa do Estado do Rio de Janeiro, Conselho Nacional de Desenvolvimento Cient\'ifico e Tecnol\'ogico and the Minist\'erio da Ci\^encia, Tecnologia e Inovac\~ao, the Deutsche Forschungsgemeinschaft, and the Collaborating Institutions in the Dark Energy Survey. The Collaborating Institutions are Argonne National Laboratory, the University of California at Santa Cruz, the University of Cambridge, Centro de Investigaciones En\'ergeticas, Medioambientales y Tecnol\'ogicas-Madrid, the University of Chicago, University College London, the DES-Brazil Consortium, the University of Edinburgh, the Eidgen\"ossische Technische Hochschule (ETH) Z\"urich, Fermi National Accelerator Laboratory, the University of Illinois at Urbana-Champaign, the Institut de Ci\`encies de l’Espai (IEEC/CSIC), the Institut de F\'isica d’Altes Energies, Lawrence Berkeley National Laboratory, the Ludwig-Maximilians Universit\"at M\"unchen and the associated Excellence Cluster Universe, the University of Michigan, NSF’s NOIRLab, the University of Nottingham, the Ohio State University, the University of Pennsylvania, the University of Portsmouth, SLAC National Accelerator Laboratory, Stanford University, the University of Sussex, and Texas A\&M University.
BASS is a key project of the Telescope Access Program (TAP), which has been funded by the National Astronomical Observatories of China, the Chinese Academy of Sciences (the Strategic Priority Research Program “The Emergence of Cosmological Structures” Grant \# XDB09000000), and the Special Fund for Astronomy from the Ministry of Finance. The BASS is also supported by the External Cooperation Program of Chinese Academy of Sciences (Grant \# 114A11KYSB20160057), and Chinese National Natural Science Foundation (Grant \# 12120101003, \# 11433005).
The Legacy Survey team makes use of data products from the Near-Earth Object Wide-field Infrared Survey Explorer (NEOWISE), which is a project of the Jet Propulsion Laboratory/California Institute of Technology. NEOWISE is funded by the National Aeronautics and Space Administration.
The Legacy Surveys imaging of the DESI footprint is supported by the Director, Office of Science, Office of High Energy Physics of the U.S. Department of Energy under Contract No. DE-AC02-05CH1123, by the National Energy Research Scientific Computing Center, a DOE Office of Science User Facility under the same contract; and by the U.S. National Science Foundation, Division of Astronomical Sciences under Contract No. AST-0950945 to NOAO.

\desied{This research used data obtained with the Dark Energy Spectroscopic Instrument (DESI). DESI construction and operations is managed by the Lawrence Berkeley National Laboratory. This material is based upon work supported by the U.S. Department of Energy, Office of Science, Office of High-Energy Physics, under Contract No. DE–AC02–05CH11231, and by the National Energy Research Scientific Computing Center, a DOE Office of Science User Facility under the same contract. Additional support for DESI was provided by the U.S. National Science Foundation (NSF), Division of Astronomical Sciences under Contract No. AST-0950945 to the NSF’s National Optical-Infrared Astronomy Research Laboratory; the Science and Technology Facilities Council of the United Kingdom; the Gordon and Betty Moore Foundation; the Heising-Simons Foundation; the French Alternative Energies and Atomic Energy Commission (CEA); the National Council of Science and Technology of Mexico (CONACYT); the Ministry of Science and Innovation of Spain (MICINN), and by the DESI Member Institutions: www.desi.lbl.gov/collaborating-institutions. The DESI collaboration is honored to be permitted to conduct scientific research on Iolkam Du’ag (Kitt Peak), a mountain with particular significance to the Tohono O’odham Nation. Any opinions, findings, and conclusions or recommendations expressed in this material are those of the author(s) and do not necessarily reflect the views of the U.S. National Science Foundation, the U.S. Department of Energy, or any of the listed funding agencies.}